# Combining Deep Learning and GARCH Models for Financial Volatility and Risk Forecasting

Jakub Michańków*, Department of Informatics, Krakow University of Economics, ul. Rakowicka 27, 31-510 Kraków; Department of Quantitative Finance, University of Warsaw, ul. Krakowskie Przedmieście 26/28, 00-927 Warszawa; e-mail: michankj@uek.krakow.pl

Łukasz Kwiatkowski, Department of Econometrics and Operations Research, Krakow University of Economics, ul. Rakowicka 27, 31-510 Kraków; e-mail: kwiatkol@uek.krakow.pl

Janusz Morajda, Department of Informatics, Krakow University of Economics, ul. Rakowicka 27, 31-510 Kraków; e-mail: morajdaj@uek.krakow.pl

* - corresponding author

**Abstract**

In this paper, we develop a hybrid approach to forecasting the volatility and risk of financial instruments by combining common econometric GARCH time series models with deep learning neural networks. For the latter, we employ Gated Recurrent Unit (GRU) networks, whereas four different specifications are used as the GARCH component: standard GARCH, EGARCH, GJR-GARCH and APARCH. Models are tested using daily logarithmic returns on the S&P 500 index as well as gold price Bitcoin prices, with the three assets representing quite distinct volatility dynamics. As the main volatility estimator, also underlying the target function of our hybrid models, we use the price-range-based Garman-Klass estimator, modified to incorporate the opening and closing prices. Volatility forecasts resulting from the hybrid models are employed to evaluate the assets' risk using the Value-at-Risk (VaR) and Expected Shortfall (ES) at two different tolerance levels of 5% and 1%. Gains from combining the GARCH and GRU approaches are discussed in the contexts of both the volatility and risk forecasts. In general, it can be concluded that the hybrid solutions produce more accurate point volatility forecasts, although it does not necessarily translate into superior VaR and ES forecasts.

# 1. Introduction and literature review

Measuring and predicting volatility and investment risk of financial assets are perennial problems of great importance both for scientists and practitioners, with the relevant literature abounding in model specifications and quantitative methods designed to address the tasks. Currently the most common approach has been developed within the area of financial econometrics, where the prices of financial instruments are typically assumed to form some conditionally heteroscedastic stochastic processes, the exact specification of which, along with their estimation and statistical inference, constitute a key part of the researchers' endeavours. A basic group of this type of tools for modelling and forecasting volatility (and, consequently, risk) are the GARCH models developed by Bollerslev (1986) & Taylor (1986), generalising the ARCH specification proposed by Engle (1982). The input information in the GARCH models, driving current volatility, comprises primarily past return rates and their conditional variances. Voluminous subsequent research aimed at modifications and extensions of the original GARCH structure, also by admitting various types of the conditional distribution. This resulted in a considerable diversity of the GARCH class, with some of the most popular specifications including EGARCH (Nelson, 1991), APARCH (Ding, Engle & Granger, 1993), GJR-GARCH (Glosten, Jagannathan & Runkle, 1993), and TGARCH (Zakoian, 1994).

A parallel trend in financial time series modelling and forecasting follows the development of machine learning tools, particularly artificial neural networks (ANNs). These models, often treated as "black boxes", are regarded as nonlinear and nonparametric techniques in which no *a priori* assumption concerning the mathematical form (equation) of the model is formulated. The function mapping input data into output signals (forecasts) is formed at the stage of training the model, implemented on the basis of a learning set including historical quotations. Over recent years, in data analyses, both researchers and practitioners have increasingly used dynamic ANNs equipped with the ability to remember and process information from some recent period of time. These tools include mainly deep-learning-based recurrent neural networks (RNNs; introduced by Hopfield, 1982, and further developed by Williams, Hinton & Rumelhart, 1986), in particular Long Short-Term Memory networks (LSTM; Hochreiter & Schmidhuber, 1997), and also (utilised in the presented research) Gated Recurrent Unit (GRU) neural networks (Chung et al., 2014), which constitute simplified modifications of LSTM.

Quite recently, a new promising research trend has emerged (including also our present paper), in which attempts are made to integrate formal tools based on the GARCH methodology with currently developed neural models based on deep learning with memorising the dynamics of the analysed phenomenon. Research on this type of hybrid models has been undertaken in many works. In particular, to cite only the most pertaining to the current paper, Kristjanpoller & Minutolo have applied hybrid models (based on feed-forward back-propagation neural network and GARCH) to predict the volatility of gold (Kristjanpoller & Minutolo, 2015) and oil prices (Kristjanpoller & Minutolo, 2016). Hu, Ni & Wen (2020) developed a hybrid deep learning method combining GARCH with neural networks and applied it to forecasting the volatility of copper price. Finally, in (Liu & So, 2020), a GARCH model was incorporated into an LSTM network for improving the prediction of stock volatility.

In this research, we analyse the predictive effectiveness of hybrid GARCH-GRU models in comparison to 'pure' GARCH models, mainly to examine the synergistic benefits of the former. The remainder of the paper is organised as follows. Section 2.1 describes methods and models used in our research, with a particular focus on developing the hybrid GARCH-GRU models. Section 2.2 depicts the data sets under study along with their statistical description and preprocessing while Section 3 presents the empirical results. Finally, Section 4 concludes.

## 2. Methodology

Below we briefly present the methodological framework of this study, the basis of which is formed by the class of GARCH models (Subsection 2.1), later combined with GRU neural networks to yield the final, GARCH-GRU hybrid specifications (Subsection 2.2). We close this section with a concise presentation of the ex post volatility forecast accuracy measures employed in our work (Subsection 2.3).

### 2.1 GARCH models

Let $r_t = 100 \cdot ln(P_t/P_{t-1})$ denote the logarithmic rate of return on some asset at time $t$, with $P_t$ and $P_{t-1}$ standing for the instrument's prices at time $t$ and $t-1$, respectively. Typically, in financial econometrics, a series of the log returns is modelled as the sum of the conditional mean of the returns (given the past information, $\psi_{t-1}$), and an error term, $\varepsilon_t$:

$$r_t = E(r_t|\psi_{t-1}) + \varepsilon_t, \tag{1}$$

where $E(r_t|\psi_{t-1})$ usually takes some ARMA form. To capture some well-documented empirical characteristics of most financial time series (with volatility clustering and fat tails among other, widely recognized features; Tsay, 2010), the errors are usually defined as some conditionally heteroscedastic process, typically of a very broad GARCH family, extending the basic ARCH structure introduced by Engle (1982), later generalised into GARCH by Bollerslev (1986). The error term therein is defined as the product:

$$\varepsilon_t = z_t h_t^{1/2}, \qquad (2)$$

where random variables $z_t \sim iiD(0, 1)$ form a sequence of independent and identically distributed standardised errors (with zero mean and unit variance), while $h_t^{1/2}$ is the return's conditional standard deviation, usually referred to as the volatility.

In this research, four most commonly entertained in the literature GARCH specifications are considered: 'standard' GARCH (Bollerslev, 1986), the Glosten-Jagannathan-Runkle GARCH (GJR-GARCH; Glosten et al. 1993), the Exponential GARCH (EGARCH; Nelson, 1991), and the Asymmetric Power ARCH (APARCH; Ding et al., 1993). Below, we briefly present their underlying equations defining the dynamics of conditional variance (a detailed and comprehensive review of univariate GARCH model specifications can be found, e.g., in Terasvirta, 2009).

*GARCH*

The volatility equation takes the form:

$$h_t = \alpha_0 + \sum_{i=1}^{q} \alpha_i \varepsilon_{t-i}^2 + \sum_{j=1}^{p} \beta_j h_{t-j}, \qquad (3)$$

where $h_t$ is the conditional variance at time $t$, and the parameters are subject to restrictions ensuring positive $h_t$ for each $t$: $\alpha_0 > 0$, $\alpha_i \geq 0$ for $i = 1, ..., q$, and $\beta_j \geq 0$ for $j = 1, ..., p$.

*GJR-GARCH*

The volatility equation takes the form:

$$h_t = \alpha_0 + \sum_{i=1}^{q} \alpha_i \varepsilon_{t-i}^2 + \sum_{i=1}^{q} \omega_i I_{t-i} \varepsilon_{t-i}^2 + \sum_{j=1}^{p} \beta_j h_{t-j}, \qquad (4)$$

where $I_{t-i} = 1$ when $\varepsilon_{t-i} \leq 0$, and $I_{t-i} = 0$ otherwise. Additionally, $\alpha_0 > 0$, $\alpha_i \geq 0$, $\alpha_i + \omega_i \geq 0$, for $i = 1, ..., q \geq 0$, and $\beta_j$ for $j = 1, ..., p$.

*EGARCH*

The volatility equation takes the form:

$$ln\ h_t = \alpha_0 + \sum_{i=1}^{q} \alpha_i\{\theta z_{t-i} + \gamma[|z_{t-i}| - E(|z_{t-i}|)]\} + \sum_{j=1}^{p} \beta_j ln\ h_{t-j}, \qquad (5)$$

where $\alpha_1 \equiv 1$ for the identification of the model.

*APARCH*

The volatility equation takes the form:

$$h^\delta_t = \alpha_0 + \sum_{i=1}^{q} \alpha_i[|\varepsilon_{t-i}| - \gamma_i \varepsilon_{t-i}]^\delta + \sum_{j=1}^{p} \beta_j h^\delta_{t-j}, \qquad (6)$$

where $\delta > 0,\ -1 < \gamma_i < 1,\ i = 1,..., q$.

Three types of conditional distributions, most commonly entertained in the literature, are used in this study for the standardised error term, $z_t$: the normal distribution, as well as Student's *t*- and skewed Student's *t*-distribution.

Estimation (through the maximum likelihood approach) and forecasting in the GARCH models have been implemented in numerous libraries available in the R programming environment, among which the `rugarch` package (see Ghalanos, 2022a, b) appears one of the most popular and comprehensive, and is also employed in this work.

## 2.2 GRU and GARCH-GRU models

The neural network component of the hybrid models developed in this paper relies on GRU neural networks, introduced by Chung et al. (2014), as a simplified version of more popular LSTM networks (see Figure 1). The GRU networks use a single unit to forget the information or update the network state, which allows them to achieve similar results to LSTMs, while significantly reducing the training time.

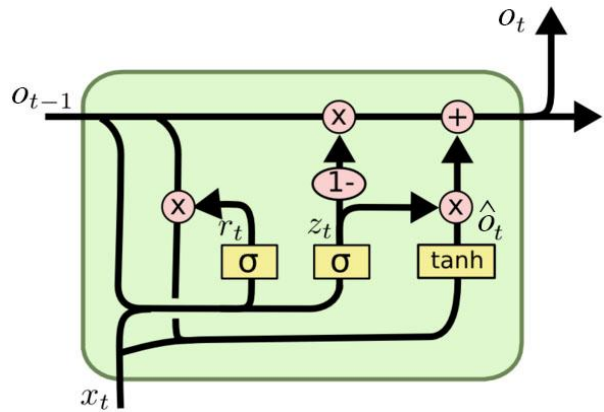

Figure 1. Architecture of a single GRU network cell
Source: Jacobs (2017), with minor modifications.

The functions within a GRU network cell can be described by the following equations:

$$z_t = \sigma_g(W_z x_t + U_z o_{t-1} + b_z), \tag{7}$$

$$r_t = \sigma_g(W_r x_t + U_r o_{t-1} + b_r), \tag{8}$$

$$\widehat{o_t} = \phi_o(W_o x_t + U_o(r_t \odot o_{t-1}) + b_o), \tag{9}$$

$$o_t = (1 - z_t) \odot o_{t-1} + z_t \odot \widehat{o_t}, \tag{10}$$

where $x_t$ is the input vector, $o_t$ is the output vector, while $z_t$ and $r_t$ are the update gate and the reset gate vectors, respectively. The matrices $W$ and $U$ (subscripted according to pertinent equations) as well as the vectors $b$ comprise the net's parameters, $\sigma_g$ and $\phi_0$ are sigmoid and hyperbolic tangent activation functions, and finally, $\odot$ denotes the Hadamard product. See Goodfellow et al. (2016) for a detailed description of the GRU networks and a comparison with other types of recurrent networks.

For the inputs, we use the following three in this research: the absolute log returns, volatility estimates (obtained by means of a given method; see below), and the volatility forecasts derived from a given GARCH model. The incorporation of the latter into the model yields a hybrid structure, further referred to as GARCH-GRU (see Figure 2).

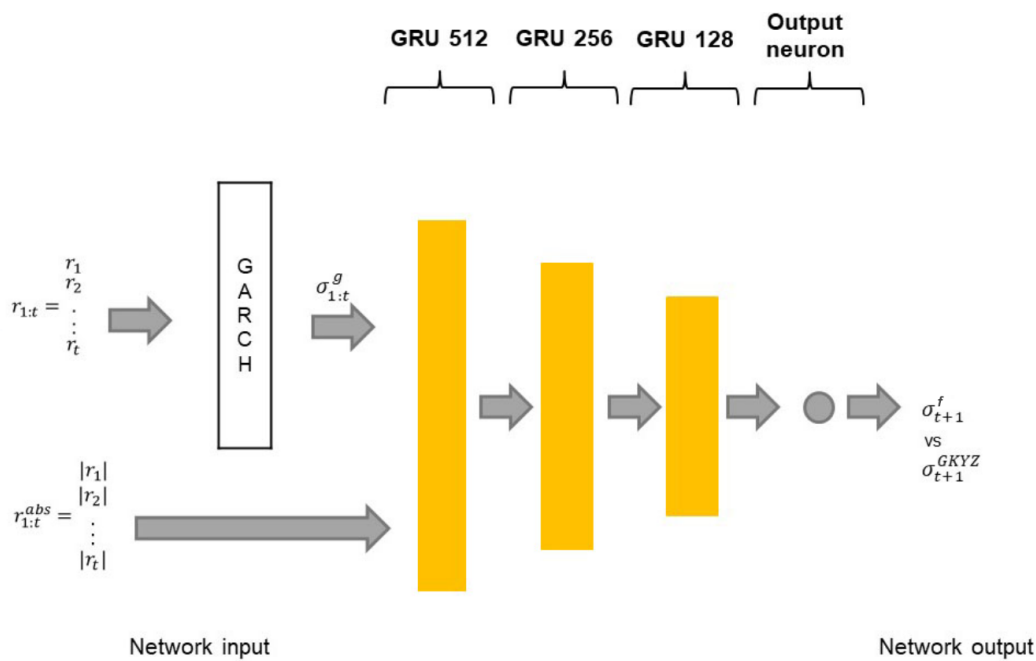

Figure 2. Hybrid GARCH-GRU model structure

Note: $\sigma_{1:t}^g$ comprises the ex post volatility forecasts from a GARCH model, $\sigma_{t+1}^{GKYZ}$ is the volatility estimated using the GKYZ estimator, and $\sigma_{t+1}^f$ is the final volatility forecast from a given hybrid model.

This model is trained to the Garman-Klass volatility estimator modified by including the gap between the previous day's closing and the current day's opening prices, following Yang & Zhang (2000). Specifically, the volatility estimate (denoted as GKYZ) is given by the formula:

$$\sigma^{2,GKYZ} = \frac{1}{n}\Sigma\left[\left(ln\frac{O_i}{C_{i-1}}\right)^2 + \frac{1}{2}\left(ln\frac{H_i}{L_i}\right)^2 - (2 \cdot ln2 - 1)\left(ln\frac{C_i}{O_i}\right)^2\right], \quad (11)$$

where $O_i$, $H_i$, $L_i$ and $C_i$, respectively, denote the opening, the highest, the lowest and the closing price at time $i$. Finally, $n$ denotes the number of daily log returns used to calculate the estimate (we set $n = 10$). Additionally, we scale the estimator to match the magnitude of the volatility estimates with the ones retrieved from GARCH models. To that end, the following formula proposed by Fiszeder (2005, 2009) is employed:

$$Scaled\ \sigma^{GKYZ} = \frac{a}{b} \cdot \sigma^{GKYZ}, \quad (12)$$

$$a = \sqrt{\frac{1}{T}\sum_{t=1}^{T} r_t^2}, \quad b = \sqrt{\frac{1}{T}\sum_{t=1}^{T} \sigma_t^{GKYZ,2}} \quad (13)$$

where $T$ denotes the initial sample size used for the estimation of a GARCH model.

Specific architecture of the GRU component used in this research consists of three GRU-type layers with 512/256/128 neurons and one single neuron dense layer on the output. Each of these GRU layers uses ReLU (Rectified Linear Unit) activation function, a dropout regulariser set to 0.3, and *l2* kernel regulariser set to 0.00001, which allows to select the best MSE value based on the validation set loss from all the epochs.

For the network optimisation, we use Adam optimiser (Kingma & Ba, 2017), with the learning rate set to 0.0009. The loss function is defined as the mean square error between the GKYZ volatility estimates at time *t*+1 and the network output (predictions; see Eq. 14). Datasets fed into the network are divided into mini-batches, with the size of 500 data points, while each batch is divided into sequences of 6 days based on which a single day prediction is produced. The tuning process is performed by means of the KersTuner with Hyperband algorithm (O'Malley et al., 2019). Finally, the model is trained for 150 epochs, with a model checkpoint callback function using the lowest value of the loss that occurred during the training.

## 2.3 Ex post volatility forecasts evaluation

Ex post assessment of the volatility forecasts produced by the various GARCH and GARCH-GRU models is carried out in this study with respect to two aspects. First, the overall point prediction accuracy is measured by three standard forecast error metrics:

$$MSE = \frac{1}{n}\sum_{t=1}^{n}(\sigma_{t+1}^{GKYZ} - \sigma_{t+1}^{f})^2, \tag{14}$$

$$MAE = \frac{1}{n}\sum_{t=1}^{n}\left|\sigma_{t+1}^{GKYZ} - \sigma_{t+1}^{f}\right|, \tag{15}$$

$$HMSE = \frac{1}{n}\sum_{t=1}^{n}\left(\frac{\sigma_{t+1}^{GKYZ} - \sigma_{t+1}^{f}}{\sigma_{t+1}^{GKYZ}}\right)^2, \tag{16}$$

where $\sigma_{t+1}^{GKYZ}$ is the volatility estimated using the GKYZ estimator and $\sigma_{t+1}^{f}$ is the volatility forecast from a given ('pure' GARCH or hybrid GARCH-GRU) model.

Differences between the MSEs for two competing models are tested for their statistical significance *via* the Diebold & Mariano (1995) test, with a modification proposed by Harvey, Leybourne & Newbold (1997). In our setting, we focus on comparing the MSE of a GARCH model with the one produced by a corresponding GARCH-GRU specification. The null hypothesis states that both of the MSE values are equal, while the alternative – that the hybrid model is more accurate than the GARCH. Finally, correlation between $\sigma_{t+1}^{2,GKYZ}$ and $\sigma_{t+1}^{2,f}$ is assessed by the coefficient of determination from the Mincer & Zarnowitz (1969) regression:

$$\sigma_{t+1}^{2,GKYZ} = \beta_0 + \beta_1 \sigma_{t+1}^{2,f} + \xi_{t+1}, \tag{17}$$

with $\xi_{t+1}$ denoting an error term.

The second aspect of the forecasts evaluation in this paper is risk measures accuracy. To that end, volatility predictions resulting from GARCH and GARCH-GRU models are used to produce long position Value at Risk (VaR) and Expected Shortfall (ES) forecasts :

$$VaR_{t+1}(\alpha) = -r_{t+1}^{f} - \sigma_{t+1}^{f} q_{\alpha}^{z}, \tag{18}$$

$$ES_{t+1}(\alpha) = E\left(r_{t+1} | r_{t+1} < VaR_{t+1}(\alpha)\right) = r_{t+1}^{f} + \sigma_{t+1}^{f} E\left(z_t | z_t < q_{\alpha}^{z}\right), \tag{19}$$

where $r^f_{t+1}$ and $\sigma^f_{t+1}$ are, respectively, the forecasted return and volatility at time *t*+1, while $q^z_\alpha$ denotes the $\alpha^{th}$ quantile of the distribution assumed for $z_t$ (see Doman & Doman, 2009; McNeil & Fray, 2000). Notice that the return predictions, $r^f_{t+1}$, are generated in our paper only through the underlying GARCH model, and thereby are not further processed through the hybrid GARCH-GRU structure. This limitation is intended here to ensure that any differences between the VaR and ES predictions stemming from GARCH and GARCH-GRU models remain attributable solely to the accuracy of the volatility forecasts produced by the two approaches. Analysis of further potential gains from using the hybrid model return predictions instead, remain beyond the scope of the current research.

Backtesting of VaR and ES forecasts is performed by means of standard tools. For the former, we use two procedures. First, the Kupiec (1995) test is employed to examine the unconditional coverage property (stated by the null hypothesis), that is the consistency between the empirical and expected numbers of VaR exceedances, under a chosen tolerance level. Both significantly higher and lower number of exceedances cause the null to be rejected. Second, through the conditional coverage test by Christoffersen (1998, 2001) test we check whether the VaR hits are independent (thus do not occur in clusters) and the empirical VaR hit ratio coincides with the assumed tolerance probability. The two statements jointly form the null hypothesis.

To backtest the Expected Shortfall predictions, we resort to the McNeil and Fray (2000) test, with the null assuming that the mean of the ES exceedances equals zero. The test results are reported in two variants, one under the exact distribution of the test statistics, while the other using a bootstrapped distribution, with the latter approach accounting for a possible misspecification of the underlying distribution of the standardised residuals.

# 3. Empirical analysis

## 3.1 Data

The following three data sets of daily logarithmic rates of return are analysed in our research, each representing quite a distinct type of financial assets: the S&P 500 index (5-day week, quotations over 6 April 2009 to 31 December 2020), Bitcoin (BTC/USD; 7-day week, quotations over 5 August 2013 to 31 December 2020), and gold (XAU/USD; 5-day week, quotations over 8 July 2010 to 31 December 2020). The time ranges of the data sets ensure an equal number of 2707 observations in each case. Figure 3 displays the asset prices, while Figure 4 shows the returns (in percentage points) and squared returns. Empirical distributions of the returns (along with a normal distribution fit) are presented in Figure 5.

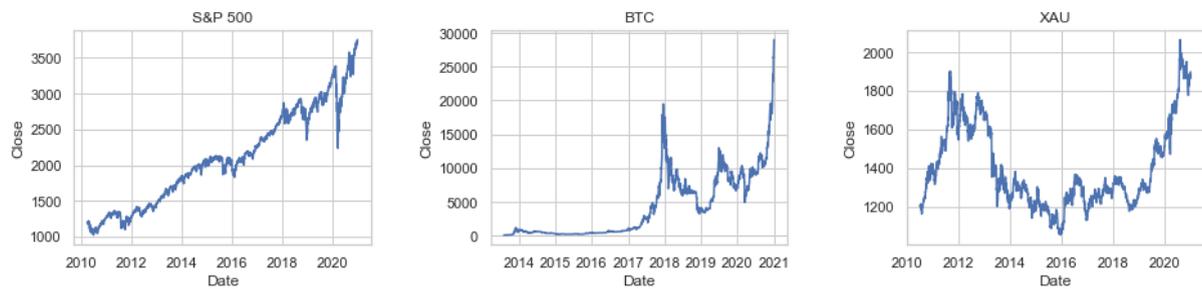

Figure 3. S&P 500, Bitcoin and gold's daily prices.

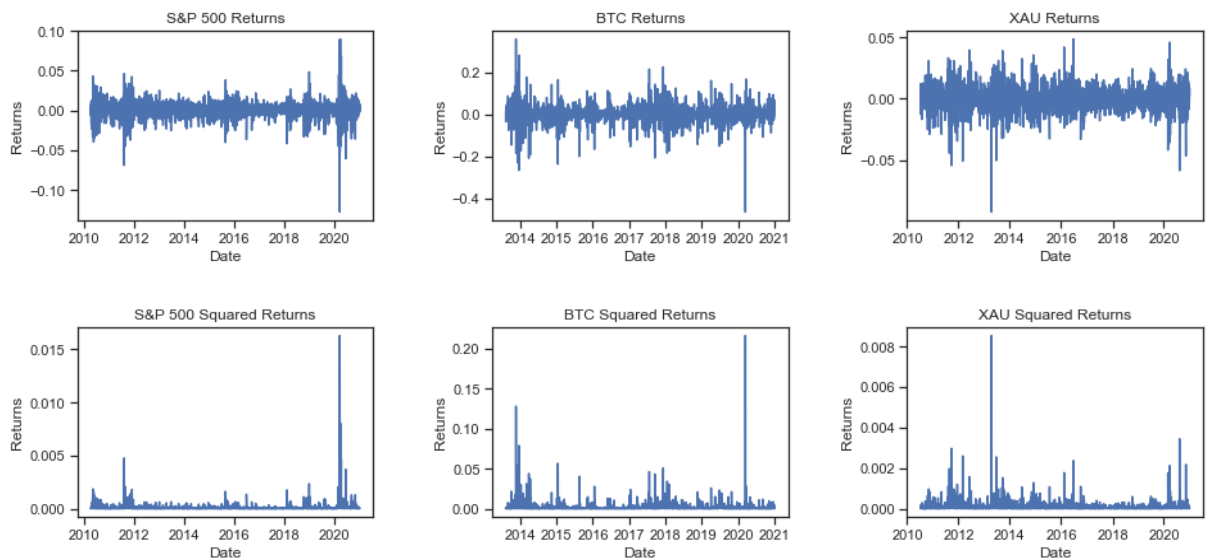

Figure 4. Returns and squared returns of S&P 500, Bitcoin and gold

Figure 4 reveals a much higher volatility of Bitcoin as compared with S&P 500 and gold, which is fairly typical, bearing in mind a generally elevated volatility of all cryptocurrencies in recent years (see the standard deviations reported in Table 1, although the coefficients of variation, CV, presented therein may suggest otherwise, which is only attributable to relatively lower means of the S&P 500 and gold returns). In addition, Bitcoin scores much higher returns (in absolute terms) than the other two assets. According to Table 1, all the data distributions exhibit a noticeable negative skewness and leptokurtosis, both quite commonly featured by financial data.

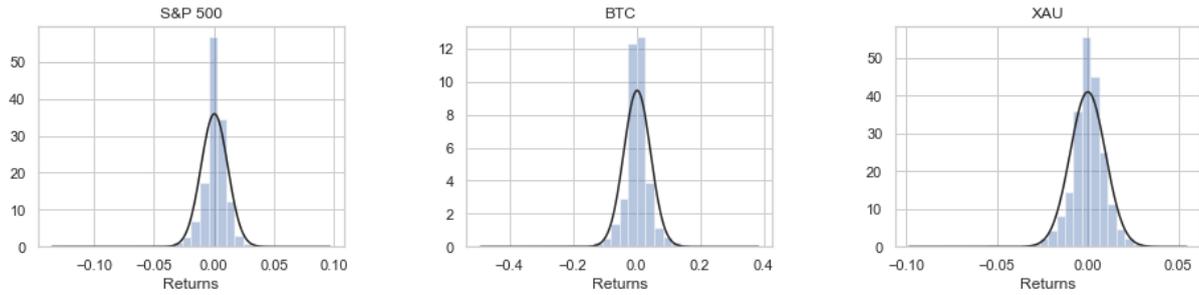

Figure 5. Histograms of S&P 500, Bitcoin and gold returns, along with a normal distribution fit (line)

Table 1. Descriptive statistics of the returns

| Statistics | S&P 500 | Bitcoin | Gold |
|---|---|---|---|
| Count | 2707 | 2707 | 2707 |
| Mean | 0.0004 | 0.0020 | 0.0001 |
| Std | 0.0111 | 0.0421 | 0.0097 |
| CV | 26.0727 | 20.2758 | 57.6982 |
| Min | -0.1276 | -0.4647 | -0.0924 |
| Max | 0.0896 | 0.3574 | 0.0486 |
| Skewness | -0.8604 | -0.5707 | -0.6435 |
| Kurtosis | 16.4489 | 12.7213 | 6.3539 |

Note: CV stands for the 'coefficient of variation.'

For the forecasting evaluation of the models (presented in the following section), all the data sets are divided in such a manner as to ensure the same amounts of observations for each asset at corresponding stages of analysis. Specifically, a rolling window scheme is employed for both GARCH and hybrid GARCH-GRU models. The size of the rolling window for the GARCH models is set to 504 days (the models are re-estimated upon arrival of each new observation). The GARCH component order is fixed to $p=q=1$ (see Eqs. 3-6) for all the assets (as typically done in the empirical literature), while the ARMA component is reduced to AR(1) for S&P 500, and only a constant for Bitcoin and gold, with the choices supported by a preliminary analysis (the results left unreported for the sake of brevity) employing the Bayesian information criterion (BIC).

For the neural network stage, the data is divided into a series of rolling training sets, each of 1008 observations, and a series of rolling test sets, each comprising 504 observations. Each time, 33% of the training set (336 observations) is used for a validation set.

The total number of the ex post evaluated predictions obtained from the GARCH and hybrid models is 1194, and is the same for each asset (although the corresponding time ranges vary: 7 April 2016 to 31 December 2020 for S&P 500, 18 May 2016 to 31 December 2020 for gold, and 29 September 2017 to 31 December 2020 for Bitcoin).

Empirical results obtained for each of the three assets are discussed below in the following fashion. First, we compare the models in terms of the MSE loss function along with testing its values through the Diebold-Mariano test, with low p-values favouring the hybrid model; see Tables 2, 5, and 8. Then, results for VaR exceedances are presented (Tables 3, 6, and 9). Finally, based on the previous, a selection of the best performing models is analysed in more detail, both with respect to the overall volatility forecast accuracy and risk prediction (Tables 4, 7, and 10), the latter including 1% and 5% Value at Risks as well as 5% Expected Shortfall.

## 3.2 Results for S&P 500

Table 2 indicates that the best performing (in terms of MSE) is the EGARCH-GRU model with a skewed Student's *t*-distribution, although only by a rather narrow margin as compared with some other specifications, like GJR-GARCH-GRU with either a symmetric or skewed *t*-distribution, and even 'standard' GARCH-GRU with a skewed *t*-distribution. Forecasts obtained from the winning specification are displayed in Figure 6, along with the GKYZ volatility estimates and the 'pure' EGARCH model (with a skewed Student's *t*-distribution).

Overall, the results presented in Table 2 imply unanimously that combining GARCH models with GRU networks significantly enhances the forecast accuracy, with all of the Diebold-Mariano test p-values remaining below 0.05.

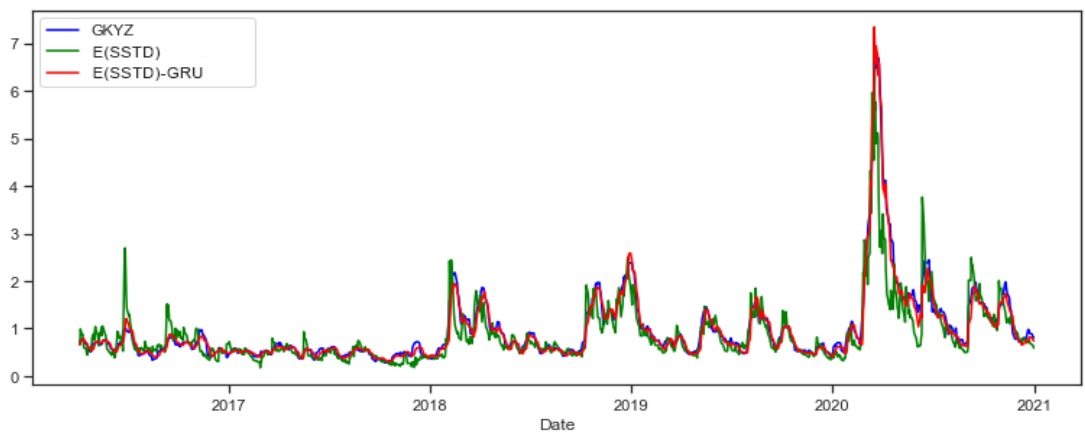

Figure 6. Volatility forecasts for S&P 500

Table 2. Comparison of volatility forecasts based on the MSE loss function across all models and distributions for S&P 500

| Metrics / Model | G(N) | G(N)-GRU | G(STD) | G(STD)-GRU | G(SSTD) | G(SSTD)-GRU |
|---|---|---|---|---|---|---|
| MSE | 0.0844 | 0.0168 | 0.0646 | 0.0183 | 0.0575 | 0.0159 |
| DM p-value | 0.0379 | | 0.0494 | | 0.0189 | |
| Metrics / Model | E(N) | E(N)-GRU | E(STD) | E(STD)-GRU | E(SSTD) | E(SSTD)-GRU |
| MSE | 0.1539 | 0.0167 | 0.1362 | 0.0160 | 0.1350 | **0.0152** |
| DM p-value | 0.0018 | | 0.0025 | | 0.0046 | |
| Metrics / Model | GJR(N) | GJR(N)-GRU | GJR(STD) | GJR(STD)-GRU | GJR(SSTD) | GJR(SSTD)-GRU |
| MSE | 0.1388 | 0.0181 | 0.1244 | 0.0153 | 0.1053 | 0.0156 |
| DM p-value | 0.0347 | | 0.0395 | | 0.0290 | |
| Metrics / Model | AP(N) | AP(N)-GRU | AP(STD) | AP(STD)-GRU | AP(SSTD) | AP(SSTD)-GRU |
| MSE | 0.1103 | 0.0333 | 0.1015 | 0.0253 | 0.0890 | 0.0221 |
| DM p-value | 0.0237 | | 0.0160 | | 0.0009 | |

Note: G stands for GARCH, E for EGARCH, GJR for GJR-GARCH, AP for APARCH. N stands for Normal distribution, STD for Student's *t*-distribution, SSTD for skewed Student's *t*-distribution. DM denotes the Diebold-Mariano test. The best result according to MSE is given in bold.

Next, we compare the models in terms of VaR unconditional coverage, with Table 3 presenting the actual number of VaR exceedances and hit ratios (in percentage terms) for both VaR tolerance levels, i.e. 5% and 1%. The results indicate that the most accurate VaR hit coverage is attained by the GJR-GARCH-GRU models with a normal and a skewed Student's *t*-distribution for the 5% tolerance level, and the APARCH model with a skewed Student's *t*-distribution for the 1% tolerance (paths of the 5% and 1% VaR forecasts along with their violations are displayed in Figure 7). Overall, and contrary to Table 2, the results here provide only mixed conclusions as to gains from resorting to the hybrid models, since combining GARCH with the GRU networks does not necessarily bring the VaR hit ratios closer to the expected 5% and 1% tolerance levels.

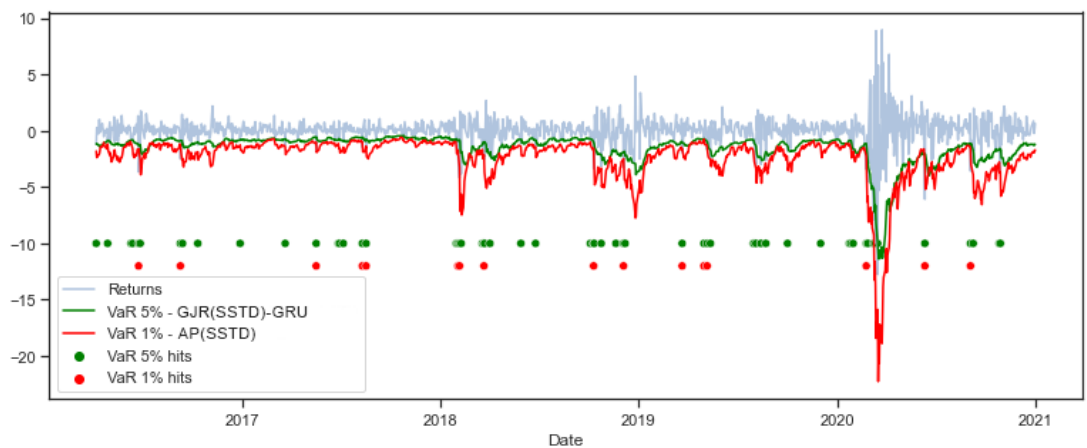

Figure 7. VaR forecasts for S&P 500

Table 3. Number of VaR exceedances across all models for S&P 500

| VaR / Model | G(N) | G(N)-GRU | G(STD) | G(STD)-GRU | G(SSTD) | G(SSTD)-GRU |
|---|---|---|---|---|---|---|
| VaR 5% | 69 (5.77%) | 62 (5.19%) | 79 (6.61%) | 72 (6.03%) | 74 (6.19%) | 63 (5.27%) |
| VaR 1% | 31 (2.59%) | 31 (2.59%) | 21 (1.75%) | 22 (1.84%) | 19 (1.59%) | 21 (1.75%) |
| VaR / Model | E(N) | E(N)-GRU | E(STD) | E(STD)-GRU | E(SSTD) | E(SSTD)-GRU |
| VaR 5% | 78 6.53%) | 58 (4.85%) | 90 (7.53%) | 63 (5.27%) | 78 (6.53%) | 56 (4.69%) |
| VaR 1% | 35 (2.93%) | 32 (2.68%) | 24 (2.01%) | 25 (2.09%) | 17 (1.42%) | 21 (1.75%) |
| VaR / Model | GJR(N) | GJR(N)-GRU | GJR(STD) | GJR(STD)-GRU | GJR(SSTD) | GJR(SSTD)-GRU |
| VaR 5% | 65 (5.44%) | **59 (4.94%)** | 75 (6.28%) | 66 (5.52%) | 68 (5.69%) | **59 (4.94%)** |
| VaR 1% | 28 (2.34%) | 34 (2.84) | 20 (1.67%) | 26 (2.17%) | 18 (1.50%) | 22 (1.84%) |
| VaR / Model | AP(N) | AP(N)-GRU | AP(STD) | AP(STD)-GRU | AP(SSTD) | AP(SSTD)-GRU |
| VaR 5% | 76 (6.36%) | 58 (6.70%) | 82 (6.86%) | 65 (5.44%) | 74 (6.19%) | 53 (4.43%) |
| VaR 1% | 34 (2.84%) | 32 (2.68%) | 22 (1.84%) | 24 (2.01%) | **16 (1.34%)** | 22 (1.84%) |

Note: G stands for GARCH, E for EGARCH, GJR for GJR-GARCH, AP for APARCH. N stands for Normal distribution, STD for Student's *t*-distribution, SSTD for skewed Student's *t*-distribution. Expected number of VaR exceedances, corresponding to 5% and 1% tolerance levels, are equal to 59 and 12, respectively. The best outcomes are indicated in bol

A selection of best performing models, according to MSE and VaR exceedances, is further analysed in more detail. Results shown in Table 4 indicate that the two GJR-GARCH-GRU models pass the unconditional coverage test for the 5% tolerance level, but rather fail the conditional coverage test, implying some clustering of the VaR violations (as might have already been expected from Figure 7, to some extent). In addition, and to one's dismay, these models also fail the ES backtest.

On the other hand, the APARCH model with a skewed Student's *t*-distribution, preferred in terms of the 1% VaR prediction, performs well in all three tests. Nevertheless, the model's performance for the 5% tolerance level is clearly surpassed by the other specifications.

Table 4. Detailed comparison of the best performing models for S&P 500

| Metrics / Model | E(SSTD)-GRU | GJR(N)-GRU | GJR(SSTD)-GRU | AP(SSTD) |
|---|---|---|---|---|
| MSE | **0.0152** | 0.0181 | 0.0156 | 0.0890 |
| MAE | **0.0787** | 0.0842 | 0.0813 | 0.1919 |
| HMSE | **0.0128** | 0.0142 | 0.0134 | 0.0550 |
| R² | 0.9736 | 0.9563 | **0.9806** | 0.8465 |
| VaR exceedances: 5%/1% | 56/21 | **59**/34 | **59**/22 | 74/**16** |
| VaR hit ratio: 5%/1% | 4.69%/1.75% | 4.94%/2.84% | 4.94%/1.84% | 6.19%/1.34% |
| Kupiec p-value: 5%/1% | 0.6197(F)/0.0173(R) | 0.9258(F)/1.6e-07(R) | 0.9258(F)/0.0088(R) | 0.0667(F)/0.2616(F) |
| Christof. p-value: 5%/1% | 0.0130(R)/0.0010(R) | 0.0017(R)/9.9e-09(R) | 0.0279(R)/0.0007(R) | 0.0555(F)/0.2408(F) |
| ES p-value bootstr./sample | 0.0551(F)/0.0155(R) | 4.9e-05(R)/1.0e-06(R) | 0.0463(R)/0.0126(R) | 0.4351(F)/0.3791(F) |

Note: VaR 5% and VaR 1% stands for 5% and 1% tolerance level. Expected number of VaR exceedances, corresponding to 5% and 1% tolerance levels, are equal to 59 and 12, respectively. F means the test failed to reject the H0 at 5% significance level, R means the H0 was rejected. G stands for GARCH model, E for EGARCH model, GJR for GJR-GARCH model, AP for APARCH model. N stands for Normal distribution, STD for Student's *t*-distribution, SSTD for skewed Student's *t*-distribution. R² is the coefficient of determination. The best (across the models) outcome according to a given metric is given in bold.

## 3.3 Results for Bitcoin

As indicated by Table 5, the best performing model for Bitcoin (in terms of point volatility forecasts) is the conditionally normal APARCH-GRU structure, with four other hybrid specifications being close seconds: GARCH-GRU and GJR-GARCH-GRU, with both symmetric and skewed *t*-distributions. As a side note, the result may imply that simpler GARCH specifications, constituting some special cases of APARCH, may require more sophisticated, heavy-tailed conditional distributions to offset their simpler volatility structure.

Overall, and similar to the case of S&P 500, combining GARCH and GRU models largely improves volatility forecasts. Enhancing a GARCH model with a GRU network results in a one- or two-order-of-magnitude drop in MSE, even though in some cases the difference appears either statistically insignificant (conditionally normal EGARCH vs. EGARCH-GRU) or at least not as statistically significant as one could expect (conditionally *t*-distributed EGARCH vs. EGARCH-GRU, and APARCH vs. APARCH-GRU with a skewed *t*-distribution). However surprising or aberrant these results may appear, they remain largely attributable to a single erratically high (compared to the target GKYZ estimates) volatility forecast obtained from the above-mentioned 'sheer' GARCH models, linked to the COVID-19 pandemic outbreak (see Figure 8). Conceivably, this volatility over-prediction is due to additional reverse transformations required to calculate the forecast of conditional standard deviation from the volatility equation defined inherently either for the logarithm of the variance (as in EGARCH; see Eq. 5) or some power transformation thereof (as in APARCH; see Eq. 6). Ultimately, these discrepancies between the 'sheer' and hybrid EGARCH (and APARCH) volatility forecasts lead to an overly high long–run variance estimate underlying the DM test, which dwindles the test statistics value, thus increasing the p-value.

Table 5. Comparison of volatility forecasts based on the MSE loss function across all models and distributions for Bitcoin

| Metrics / Model | G(N) | G(N)-GRU | G(STD) | G(STD)-GRU | G(SSTD) | G(SSTD)-GRU |
|---|---|---|---|---|---|---|
| MSE | 5.6795 | 0.4042 | 5.0206 | 0.3938 | 5.0327 | 0.3937 |
| DM p-value | 2.537e-05 | | 4.399e-16 | | 4.233e-16 | |
| Metrics / Model | E(N) | E(N)-GRU | E(STD) | E(STD)-GRU | E(SSTD) | E(SSTD)-GRU |
| MSE | 7.4872 | 0.4406 | 11.3832 | 0.4698 | 62.5108 | 0.5980 |
| DM p-value | 0.1355 | | 0.006316 | | 1.429e-13 | |
| Metrics / Model | GJR(N) | GJR(N)-GRU | GJR(STD) | GJR(STD)-GRU | GJR(SSTD) | GJR(SSTD)-GRU |
| MSE | 5.8335 | 0.4202 | 5.0074 | 0.3946 | 5.0188 | 0.3982 |
| DM p-value | 5.838e-05 | | 1.844e-15 | | 1.8e-15 | |
| Metrics / Model | AP(N) | AP(N)-GRU | AP(STD) | AP(STD)-GRU | AP(SSTD) | AP(SSTD)-GRU |
| MSE | 5.8356 | **0.3818** | 3.7345 | 0.4308 | 12.9127 | 0.4184 |
| DM p-value | 0.000837 | | 3.167e-08 | | 0.01944 | |

Note: G stands for GARCH, E for EGARCH, GJR for GJR-GARCH, AP for APARCH. N stands for Normal distribution, STD for Student's *t*-distribution, SSTD for skewed Student's *t*-distribution. DM denotes the Diebold-Mariano test. The best result according to MSE is given in bold.

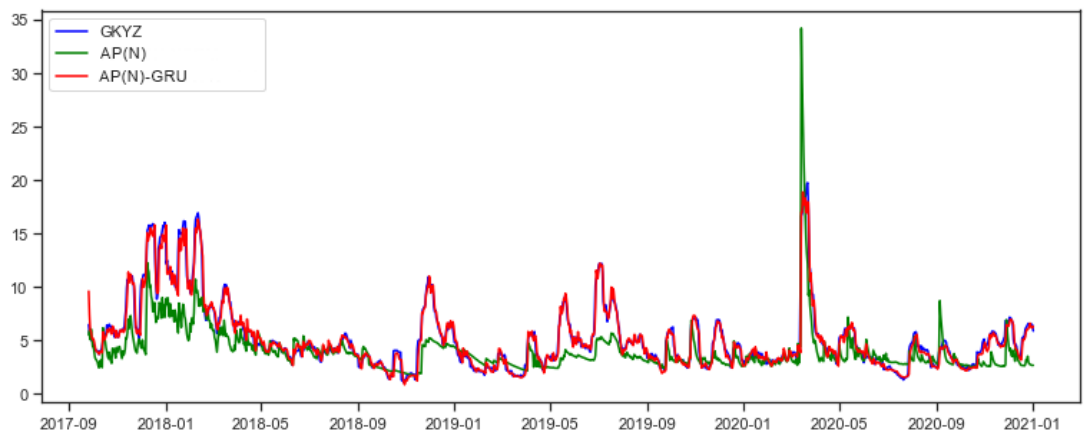

Figure 8. Volatility forecasts for Bitcoin

Table 6 presents the number and hit ratios of VaR exceedances. Despite the earlier results indicating a considerable gain from combining GARCH with GRU models for the sake of volatility forecasting, the effect does not necessarily translate into superior VaR prediction performance of the hybrid structures, to one's dismay. For both of the tolerance levels under consideration, it is a 'sheer'

APARCH model that produces VaR estimates with a hit ratio nearest to the expected one: the conditionally normal APARCH for the 5% tolerance level, and APARCH with a *t*-distribution for the 1% tolerance (see Figure 9). Overall, as inferred from Table 6, combining GARCH with GRU models may lead to either more conservative or more liberal VaR predictions, as compared with the ones from the corresponding GARCH structures.

Table 6. Number of VaR exceedances across all models for Bitcoin

| VaR / Model | G(N) | G(N)-GRU | G(STD) | G(STD)-GRU | G(SSTD) | G(SSTD)-GRU |
|---|---|---|---|---|---|---|
| VaR 5% | 52 (4.35%) | 33 (2.76%) | 70 (5.86%) | 50 (4.18%) | 69 (5.77%) | 46 (3.85%) |
| VaR 1% | 23 (1.92%) | 14 (1.17%) | 16 (1.34%) | 13 (1.08%) | 16 (1.34%) | 14 (1.17%) |
| VaR / Model | E(N) | E(N)-GRU | E(STD) | E(STD)-GRU | E(SSTD) | E(SSTD)-GRU |
| VaR 5% | 51 (4.27%) | 32 (2.68%) | 63 (5.27%) | 81 (6.78%) | 64 (5.36%) | 94 (7.87%) |
| VaR 1% | 23 (1.92%) | 13 (1.08%) | 13 (1.08%) | 19 (1.59%) | 11 (0.92%) | 33 (2.76%) |
| VaR / Model | GJR(N) | GJR(N)-GRU | GJR(STD) | GJR(STD)-GRU | GJR(SSTD) | GJR(SSTD)-GRU |
| VaR 5% | 54 (4.52%) | 33 (2.76%) | 71 (5.94%) | 47 (3.93%) | 67 (5.61%) | 43 (3.60%) |
| VaR 1% | 24 (2.01%) | 14 (1.17%) | 16 (1.34%) | 13 (1.08%) | 16 (1.34%) | 13 (1.08%) |
| VaR / Model | AP(N) | AP(N)-GRU | AP(STD) | AP(STD)-GRU | AP(SSTD) | AP(SSTD)-GRU |
| VaR 5% | **56 (4.69%)** | 32 (2.68%) | 63 (5.27%) | 67 (5.61%) | 55 (4.60%) | 65 (5.44%) |
| VaR 1% | 23 (1.92%) | 14 (1.17%) | **12 (1.00%)** | 14 (1.17%) | 11 (0.92%) | 18 (1.50%) |

Note: G stands for GARCH, E for EGARCH, GJR for GJR-GARCH, AP for APARCH. N stands for Normal distribution, STD for Student's *t*-distribution, SSTD for skewed Student's *t*-distribution. Expected number of VaR exceedances, corresponding to 5% and 1% tolerance levels, are equal to 59 and 12, respectively. The best outcomes are indicated in bold.

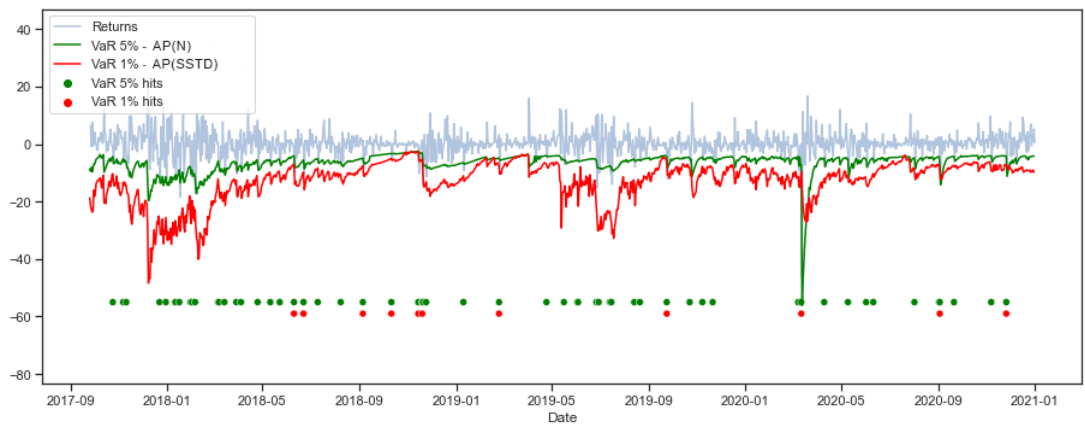

Figure 9. VaR forecasts for Bitcoin

Table 7 presents detailed results for a selection of the best models (based on the MSE and VaR coverage performance). In general, the discrepancy between the models' performance in terms of either volatility forecasting or VaR and ES prediction is striking. The conditionally normal APARCH model combined with a GRU network produces by far superior forecasts of the returns' volatility, yet fails to yield unanimously satisfactory outcomes for risk prediction. Conversely, it is 'sheer' APARCH models with a normal and Student's *t*-distribution that yield the best VaR and ES forecasts for 5% and 1% tolerances, respectively (however poorly forecasting the volatility itself). Such a divergence of the models' performance may lead one to a conclusion that the GKYZ volatility estimates, employed in training the GRU component of the hybrid model structures, are somewhat deficient for the follow-up task of risk assessment. Conceivably, this may be the case due to not only a generally high volatility of the Bitcoin returns, but also a strikingly high 'volatility of volatility', numerous and pronounced spikes in the modelled series, interspersing otherwise relatively 'regular' returns. (see Figure 4).

Table 7. Detailed comparison of the best performing models for Bitcoin

| Metrics / Model | APARCH(N)-GRU | APARCH(N) | APARCH(STD) |
| --- | --- | --- | --- |
| MSE | **0.3818** | 5.835 | 3.7345 |
| MAE | **0.3906** | 1.5928 | 1.3160 |
| HMSE | **0.0145** | 0.1024 | 0.1290 |
| $R^2$ | **0.9586** | 0.3186 | 0.5928 |
| VaR exceedances: 5%/1% | 32/14 | **56**/23 | 63/**12** |
| VaR hit ratio: 5%/1% | 2.68%/1.17% | **4.69%**/1.92% | 5.36%/**1%** |
| Kupiec p-value: 5%/1% | 5.81e-05(R) / 0.5597(F) | 0.6197(F) / 0.0043(R) | 0.6639(F) / 0.9860(F) |
| Christof. p-value: 5%/1% | 0.0003(R) / 0.7143(F) | 0.6265(F) / 0.0108(R) | 0.1548(F) / 0.8850(F) |
| ES p-value bootstr./sample | 0.0368(R) / 0.0360(R) | 0.0118(R) / 0.0097(R) | 0.7616(F) / 0.8199(F) |

Note: VaR 5% and VaR 1% stands for 5% and 1% tolerance level. Expected number of VaR exceedances, corresponding to 5% and 1% tolerance levels, are equal to 59 and 12, respectively. F means the test failed to reject the H0 at 5% significance level, R means the H0 was rejected. G stands for GARCH model, E for EGARCH model, GJR for GJR-GARCH model, AP for APARCH model. N stands for Normal distribution, STD for Student's *t*-distribution, SSTD for skewed Student's *t*-distribution. $R^2$ is a coefficient of determination. The best (across the models) outcome according to a given metric is given in bold.

### 3.4 Results for gold

According to Table 8, the most accurate volatility forecasts for the returns on gold are produced by a hybrid EGARCH-GRU model with a Student's *t*-distribution (see Figure 10), although most of the remaining hybrid specifications could be regarded as 'close seconds'.

Overall, combining the GARCH with GRU models largely enhances the prediction throughout all the resulting specifications. However, similarly to the two previous two assets, the 'hybridization' does not improve the VaR prediction unanimously (see Table 9). Nevertheless, it is still the hybrid models: *t*-distributed GARCH-GRU, normally distributed EGARCH-GRU, and GJR-GRACH-GRU with a skewed *t*-distribution, that perform the best at the 5% tolerance level. On the contrary, 'sheer' GARCH, EGARCH and GJR-GARCH models (with a skewed *t*-distribution) prove the best for the 1% VaR prediction.

Table 8. Comparison of volatility forecasts based on the MSE loss function across all models and distributions for gold

| Metrics / Model | G(N) | G(N)-GRU | G(STD) | G(STD)-GRU | G(SSTD) | G(SSTD)-GRU |
|---|---|---|---|---|---|---|
| MSE | 0.0376 | 0.0078 | 0.037 | 0.0090 | 0.0378 | 0.0078 |
| DM p-value | 1.62e-06 | | 1.39e-06 | | 7.5e-07 | |
| Metrics / Model | E(N) | E(N)-GRU | E(STD) | E(STD)-GRU | E(SSTD) | E(SSTD)-GRU |
| MSE | 0.0469 | 0.0078 | 0.0485 | **0.0076** | 0.0493 | 0.0081 |
| DM p-value | 0.0001 | | 0.0001 | | 0.0001 | |
| Metrics / Model | GJR(N) | GJR(N)-GRU | GJR(STD) | GJR(STD)-GRU | GJR(SSTD) | GJR(SSTD)-GRU |
| MSE | 0.0450 | 0.0079 | 0.0505 | 0.0083 | 0.0510 | 0.0083 |
| DM p-value | 6.55e-05 | | 1.16e-05 | | 1.52e-05 | |
| Metrics / Model | AP(N) | AP(N)-GRU | AP(STD) | AP(STD)-GRU | AP(SSTD) | AP(SSTD)-GRU |
| MSE | 0.0441 | 0.0080 | 0.0453 | 0.0079 | 0.0457 | 0.0079 |
| DM p-value | 0.0007 | | 5.76e-05 | | 8.63e-05 | |

Note: G stands for GARCH, E for EGARCH, GJR for GJR-GARCH, AP for APARCH. N stands for Normal distribution, STD for Student's *t*-distribution, SSTD for skewed Student's *t*-distribution. DM denotes the Diebold-Mariano test. The best result according to MSE is given in bold.

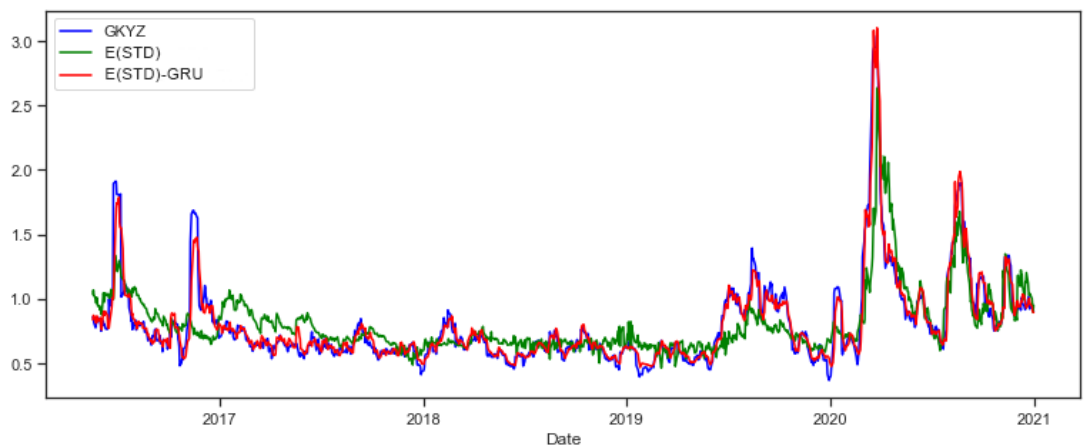

Figure 10. Volatility forecasts for gold

Table 9. Number of VaR exceedances across all models for gold

| VaR / Model | G(N) | G(N)-GRU | G(STD) | G(STD)-GRU | G(SSTD) | G(SSTD)-GRU |
|---|---|---|---|---|---|---|
| VaR 5% | 64 (5.36%) | 57 (4.77%) | 72 (6.03%) | **59 (4.94%)** | 68 (5.69%) | 61 (5.10%) |
| VaR 1% | 21 (1.75%) | 17 (1.42%) | 15 (1.25%) | 9 (0.75%) | **13 (1.08%)** | 10 (0.83%) |
| VaR / Model | E(N) | E(N)-GRU | E(STD) | E(STD)-GRU | E(SSTD) | E(SSTD)-GRU |
| VaR 5% | 60 (5.02%) | **59 (4.94%)** | 64 (5.36%) | 62 (5.19%) | 61 (5.10%) | 61 (5.10%) |
| VaR 1% | 25 (2.09%) | 19 (1.59%) | 17 (1.42%) | 9 (0.75%) | **13 (1.08%)** | 10 (0.83%) |
| VaR / Model | GJR(N) | GJR(N)-GRU | GJR(STD) | GJR(STD)-GRU | GJR(SSTD) | GJR(SSTD)-GRU |
| VaR 5% | 63 (5.27%) | 58 (4.85%) | 67 (5.61%) | 62 (5.19%) | 62 (5.19%) | **59 (4.94%)** |
| VaR 1% | 26 (2.17%) | 19 (1.59%) | 14 (1.17%) | 9 (0.75%) | **13 (1.08%)** | 10 (0.83%) |
| VaR / Model | AP(N) | AP(N)-GRU | AP(STD) | AP(STD)-GRU | AP(SSTD) | AP(SSTD)-GRU |
| VaR 5% | 66 (5.52%) | 60 (5.02%) | 69 (5.77%) | 64 (5.36%) | 68 (5.69%) | 61 (5.10%) |
| VaR 1% | 28 (2.34%) | 19 (1.59%) | 17 (1.42%) | 9 (0.75%) | 16 (1.34%) | 14 (1.17%) |

Note: G stands for GARCH, E for EGARCH, GJR for GJR-GARCH, AP for APARCH. N stands for Normal distribution, STD for Student's *t*-distribution, SSTD for skewed Student's *t*-distribution. Expected number of VaR exceedances, corresponding to 5% and 1% tolerance levels, are equal to 59 and 12, respectively. The best outcomes are indicated in bold.

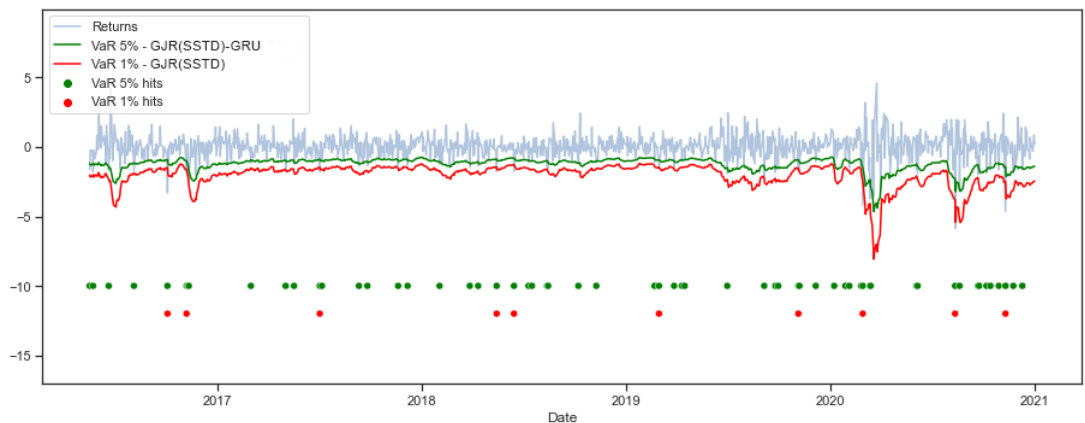

Figure 11. VaR forecasts for gold

Table 10 presents a detailed analysis for a selection of the best performing models for gold. The results are fairly mixed, with indications of a superior specification hinged on the choice of a particular metrics, thereby precluding an ultimate winner. Nevertheless, they still partially support the hybrid approach of combining the GARCH with GRU models.

Table 10. Detailed comparison of the best performing models for gold

| Metrics / Model | E(STD)-GRU | GJR(SSTD)-GRU | GJR(SSTD) | E(N)-GRU |
|---|---|---|---|---|
| MSE | **0.0076** | 0.0083 | 0.0510 | 0.00781 |
| MAE | 0.0577 | 0.0588 | 0.1609 | **0.0557** |
| HMSE | 0.0107 | 0.0108 | 0.0680 | **0.0097** |
| $R^2$ | **0.9417** | 0.9288 | 0.5357 | 0.9356 |
| VaR exceedances: 5%/1% | 62/9 | **59**/9 | 68/**13** | **59**/19 |
| VaR hit ratio: 5%/1% | 5.19%/0.75% | **4.94%**/0.75% | 5.69%/**1.08%** | **4.94%**/1.59% |
| Kupiec p-value: 5%/1% | 0.7614(F)/0.3713(F) | 0.9258(F)/0.5615(F) | 0.7614(F)/0.7611(F) | 0.9258(F)/0.058(F) |
| Christof. p-value: 5%/1% | 0.9500(F)/0.6262(F) | 0.8475(F)/0.7764(F) | 0.8534(F)/0.8274(F) | 0.8475(F)/0.1230(F) |
| ES p-value bootstr./sample | 0.4312(F)/0.3691(F) | 0.4108(F)/0.3517(F) | 0.2453(F)/0.1624(F) | 0.0264(R)/0.0109(R) |

Note: VaR 5% and VaR 1% stands for 5% and 1% tolerance level. Expected number of VaR exceedances, corresponding to 5% and 1% tolerance levels, are equal to 59 and 12, respectively. F means the test failed to reject the H0 at 5% significance level, R means the H0 was rejected. G stands for GARCH model, E for EGARCH model, GJR for GJR-GARCH model, AP for APARCH model. N stands for Normal distribution, STD for Student's *t*-distribution, SSTD for Skewed Student's *t*-distribution. $R^2$ is a coefficient of determination. The best (across the models) outcome according to a given metric is given in bold.

# 4. Conclusions

The main aim of the paper was to develop hybrid GARCH-GRU models for the task of financial volatility and risk prediction, thereby bridging the most common, 'classic' econometric tools for volatility dynamics (GARCH models) with deep machine learning methods. The approach was tested on three financial assets displaying distinct volatility dynamics: S&P 500, Bitcoin and gold.

In summary, it can be concluded that no single model specification would do best in all the situations. Overall, however, for each of the three assets under consideration, it was a hybrid model that emerged superior for the point volatility forecasting (in terms of MSE): EGARCH-GRU models for S&500 and gold (under a skewed and a symmetric *t*-distribution, respectively), while APARCH-GRU (with a normal distribution) – for Bitcoin. Nonetheless, this general outcome is hardly a surprise, given that the main task of the GRU network in the hybrid models was to minimise the MSE loss.

Somewhat to one's dismay, the gains from the volatility prediction by means of the hybrid GARCH-GRU structures do not appear to translate unanimously into superior Value at Risk and Expected Shortfall forecasts. From Tables 4, 7 and 10 it can be inferred that the choice of a winning specification largely hinges on both the asset at hand as well as tolerance level. Using, for brevity, the models' acronyms used in the tables, the following models proved the most valid with respect to the risk assessment at the tolerance levels of 5% and 1%:

- S&P 500: GJR(SSTD)-GRU at the 5% tolerance, and AP(SSTD) at 1%,
- Bitcoin: AP(N) at 5%, and AP(STD) at 1%,
- gold: G(STD)-GRU and E(N)-GRU at 5%, and G(SSTD) at 1%.

The above list indicates clearly that hybridising GARCH with GRU models does not necessarily yield superior risk forecasts. Moreover, all of the listed specifications differ from the ones that proved most accurate for volatility forecasting in terms of MSE, mentioned in the previous paragraph. This, in turn, may actually put into question the very choice of the target function underlying the GRU components in the hybrid models advanced in this paper (here hinged on the point volatility forecast accuracy), therefore necessitating the function to be redefined specifically for the task of VaR and/or ES prediction. This line of research is left for future work.

On the whole, the research findings corroborate the potential and purposefulness of combining 'classic' econometric models for volatility dynamics with deep machine learning approaches for the purpose of improving, in general, results produced by the former. Nevertheless, the results presented in the current paper preclude unanimous conclusions as to the empirical advantages of such 'hybridisation', leaving it largely to a particular financial asset and task at hand.

**Acknowledgements**

The authors acknowledge financial support from subsidies granted to the Krakow University of Economics.

# References

1. Aloui, C., Ben Hamida, H. (2015). Estimation and Performance Assessment of Value-at-Risk and Expected Shortfall Based on Long-Memory GARCH-Class Models. Finance a Uver: *Czech Journal of Economics & Finance,* Vol. 65: 30–54.
2. Aloui C., Mabrouk S. (2010). Value-at-risk estimations of energy commodities via long-memory, asymmetry and fat-tailed GARCH models. *Energy Policy*, *Greater China Energy: Special Section with regular papers*, Vol. 38: 2326–2339.
3. Ardia D., Hoogerheide L.F. (2014). GARCH models for daily stock returns: Impact of estimation frequency on Value-at-Risk and Expected Shortfall forecasts. *Economics Letters*, Vol. 123: 187–190.

4. Bollerslev T. (1986), Generalized Autoregressive Conditional Heteroskedasticity, *Journal of Econometrics*, Vol 31 (3): 307–3
5. Bams, D., Blanchard, G., Lehnert, T. (2017). Volatility measures and Value-at-Risk. *International Journal of Forecasting,* Vol. 33: 848–863.
6. Cho, K., van Merrienboer, B., Gulcehre, C., Bahdanau, D., Bougares, F., Schwenk, H., Bengio, Y. (2014). Learning Phrase Representations using RNN Encoder-Decoder for Statistical Machine Translation. arXiv:1406.1078 [Cs, Stat]. http://arxiv.org/abs/1406.1078, Accessed 22 August 2023.
7. Chung, J., Gulcehre, C., Cho, K., Bengio, Y. (2014). *Empirical Evaluation of Gated Recurrent Neural Networks on Sequence Modeling*. arXiv:1412.3555 [cs.NE], https://arxiv.org/abs/1412.3555, Accessed 22 August 2023.
8. Christoffersen, P. F. (1998). Evaluating Interval Forecasts. *International Economic Review*, Vol: 39(4): 841–862.
9. Christoffersen, P., Hahn, J., Inoue, A. (2001). Testing and Comparing Value-at-Risk Measures. *Journal of Empirical Finance* Vol. 8(3): 325-342.
10. Diebold, F. X., Mariano, R. S. (1995). Comparing Predictive Accuracy. *Journal of Business & Economic Statistics*, Vol. 13(3): 253–263.
11. Ding Z., Engle R.F., Granger C.W.J. (1993), A long memory property of stock market return and a new model, *Journal of Empirical Finance,* Vol. 1(1): 83-106,
12. Doman M., Doman R. (2009). *Modelowanie zmienności i ryzyka. Metody ekonometrii finansowej*, Wolters Kluwer, Kraków.
13. Du, Z., Wang, M., & Xu, Z. (2019). *On Estimation of Value-at-Risk with Recurrent Neural Network*, 2019 Second International Conference on Artificial Intelligence for Industries (AI4I), 103–106.
14. Engle R.F. (1982), Autoregressive Conditional Heteroscedasticity with Estimates of Variance of United Kingdom Inflation, *Econometrica*, Vol. 50 (4): 987–1008.
15. Fischer, T., & Krauss, C. (2018). Deep learning with long short-term memory networks for financial market predictions. *European Journal of Operational Research*, Vol. 270(2), 654–669.
16. Fiszeder P. (2005), Forecasting the volatility of the Polish stock index – WIG20. W: *Forecasting Financial Markets. Theory and Applications*. Łódź.
17. Fiszeder P. (2007), Prognozowanie VaR – zastosowanie wielorównaniowych modeli GARCH, Modelowanie i prognozowanie gospodarki narodowej, *Prace i Materiały Wydziału Zarządzania Uniwersytetu Gdańskiego*, 365-376.
18. Fiszeder P. (2009), *Modele klasy GARCH w empirycznych badaniach finansowych*, Wydawnictwo Naukowe Uniwersytetu Mikołaja Kopernika, Toruń.
19. Garman, M. B., & Klass, M. J. (1980). On the Estimation of Security Price Volatilities from Historical Data. *The Journal of Business*, Vol. 53(1): 67–78.
20. Ghalanos, A. (2022a). *Introduction to the rugarch package* (Version 1.4-3), https://cran.r-project.org/web/packages/rugarch/vignettes/Introduction_to_the_rugarch_package.pdf, Accessed 22 August 2023.

21. Ghalanos, A. (2022b). *rugarch: Univariate GARCH models*, https://cran.r-project.org/web/packages/rugarch/rugarch.pdf, Accessed 22 August 2023.
22. Glosten L.R., Jagannathan R., Runkle D.E. (1993), Relationship between the expected value and the volatility of the nominal excess return on stocks, *The Journal of Finance*, Vol. 48(5): 1779-1801.
23. Goodfellow I., Bengio Y., Courville A. (2016). *Deep Learning*, The MIT Press.
24. Harvey, D., Leybourne, S., Newbold, P. (1997). Testing the equality of prediction mean squared errors. *International Journal of forecasting*, Vol. 13(2): 281-291
25. Higgins M.L., Bera A.K. (1992), *A class of nonlinear ARCH models*, International Economic Review, Vol: 33 (1): 137-158.
26. Hochreiter S., Schmidhuber J. (1997), *Long Short-Term Memory*. Neural Computation. Vol. 9(8): 1735–1780.
27. Hopfield J.J. (1982), *Neural networks and physical systems with emergent collective computational abilities*, Proceedings of the National Academy of Sciences, Vol. 79(8): 2554–2558.
28. Hu Y., Ni J., Wen L. (2020), A hybrid deep learning approach by integrating LSTM-ANN networks with GARCH model for copper price volatility prediction, *Physica A: Statistical Mechanics and its Applications*, Vol. 557, Article 124907.
29. Jacobs, K. (2017), *Python Deep Learning Tutorial: Create A GRU (RNN) In TensorFlow,* https://www.data-blogger.com/2017/08/27/gru-implementation-tensorflow/, Accessed 10 December 2021.
30. Kim, H. Y., Won, C. H. (2018). Forecasting the volatility of stock price index: A hybrid model integrating LSTM with multiple GARCH-type models. *Expert Systems with Applications*, Vol. 103: 25–37.
31. Kingma, D. P., Ba, J. (2017). *Adam: A Method for Stochastic Optimization.* ArXiv:1412.6980 [Cs], http://arxiv.org/abs/1412.6980, Accessed 22 August 2023.
32. Kristjanpoller, W., & Hernández, E. (2017). Volatility of main metals forecasted by a hybrid ANN-GARCH model with regressors. *Expert Systems with Applications*, Vol. 84: 290–300.
33. Kristjanpoller, W., & Minutolo, M. C. (2018). A hybrid volatility forecasting framework integrating GARCH, artificial neural network, technical analysis and principal components analysis. *Expert Systems with Applications*, Vol. 109: 1–11.
34. Kristjanpoller W., Minutolo M.C. (2016), Forecasting volatility of oil price using an artificial neural network-GARCH model, *Expert Systems with Applications*, Vol. 65: 233-241,
35. Kristjanpoller W., Minutolo M.C. (2015), Gold price volatility: A forecasting approach using the Artificial Neural Network–GARCH model, *Expert Systems with Applications*, Vol. 42: 7245-7251.
36. Kupiec, P. (1995). *Techniques for Verifying the Accuracy of Risk Measurement Models* (SSRN Scholarly Paper ID 6697). Social Science Research Network, https://papers.ssrn.com/abstract=6697, Accessed 22 August 2023.

37. Liu W.K., & So M.K.P. (2020), A GARCH model with artificial neural networks, *Information*, Vol. 11(10): 489.
38. McNeil, A. J., & Frey, R. (2000), Estimation of tail-related risk measures for heteroscedastic financial time series: an extreme value approach. *Journal of Empirical Finance*, Vol. *7*(3): 271–300.
39. Małecka, M. (2016), *Weryfikacja hipotez w ocenie ryzyka rynkowego*, Wydawnictwo Uniwersytetu Łódzkiego, Łódź.
40. Mincer, J., & Zarnowitz, V., (1969), *The Evaluation of Economic Forecasts*. In: Economic Forecasts and Expectations: Analysis of Forecasting Behavior and Performance, National Bureau of Economic Research, Inc, https://EconPapers.repec.org/RePEc:nbr:nberch:1214.
41. Nelson D.B. (1991), Conditional heteroscedasticity in asset returns: a new approach, *Econometrica*, Vol. 59(2): 347-370.
42. O'Malley, T., Bursztein, E., Long, J., Chollet, F., Jin, H., Invernizzi, L. and others, (2019), *KerasTuner*. https://github.com/keras-team/keras-tuner, Accessed 22 August 2023.
43. Piontek K., Papla D. (2005), Wykorzystanie wielorównaniowych modeli AR-GARCH w pomiarze ryzyka metodą VaR, *Prace Naukowe Akademii Ekonomicznej we Wrocławiu*, s. 126-138
44. Taylor S. (1986), *Modelling Financial Time Series*. Wiley.
45. Teräsvirta, T. (2009). An Introduction to Univariate GARCH Models. In: Mikosch, T., Kreiß, JP., Davis, R., Andersen, T. (eds) *Handbook of Financial Time Series*. Springer, Berlin, Heidelberg.
46. Tsay R.S. (2010), *Analysis of Financial Time Series*, John Wiley & Sons, Chicago.
47. Williams R.J., Hinton G.E., Rumelhart D.E. (1986), Learning representations by back-propagating errors, *Nature*, Vol. 323 (6088): 533–536.
48. Yang, D., & Zhang, Q. (2000), Drift‑Independent Volatility Estimation Based on High, Low, Open, and Close Prices. *The Journal of Business*, Vol. 73(3): 477–492.
49. Zakoian J.M. (1994), Threshold heteroscedasticity models. *Journal of Economic Dynamics and Control*, Vol. 18(5): 931-955.